**Trustworthiness of statistical inference**

David J. Hand
Department of Mathematics, Imperial College, London SW7 2AZ
d.j.hand@imperial.ac.uk

## Summary

We examine the role of trustworthiness and trust in statistical inference, arguing that it is the extent of trustworthiness in inferential statistical tools which enables trust in the conclusions. Certain tools, such as the p-value and significance test, have recently come under renewed criticism, with some arguing that they damage trust in statistics. We argue the contrary, beginning from the position that the central role of these methods is to form the basis for trusted conclusions in the face of uncertainty in the data, and noting that it is the misuse and misunderstanding of these tools which damages trustworthiness and hence trust. We go on to argue that recent calls to ban these tools would tackle the symptom, not the cause, and themselves risk damaging the capability of science to advance, and feeding into public suspicion of the discipline of statistics. The consequence could be aggravated mistrust of our discipline and of science more generally. In short, the very proposals could work in quite the contrary direction from that intended. We make some alternative proposals for tackling the misuse and misunderstanding of these methods, and for how trust in our discipline might be promoted.

## 1. Introduction

Recent years have seen a growth in interest in the concepts of trust and trustworthiness. In part this might be a reaction to notions such as "false facts" and "fake news", but interest and public attention were growing before these became mainstream. For example, in 1998 the UK government launched a consultation entitled *Statistics: A Matter of Trust*, with the foreword saying "official statistics must above all be trustworthy and be seen to be trustworthy" (UK Government, 1998), and in 2002 Onora O'Neill's Reith Lectures, *A Question of Trust*, were broadcast by the BBC (O'Neill, 2002). More recently, in a statistical context, we have seen the UK Statistics Authority adopting trustworthiness as one of the three pillars underpinning its revised Code of Practice (UKSA, 2018) (the other two being quality and value) and elsewhere in official statistics the OECD is funding "innovative projects that enhance trust in data and statistics in low and middle-income countries" via its *PARIS21 2020 Trust Initiative* (PARIS21, 2020). Yet another official statistics strand is the notion of "trusted smart statistics" (see Ricciato *et al*, 2019; Vichi and Hand 2019; and the other seven papers on this topic in a special issue of the *Journal of the IOAS*, Vol.35:4). Trusted smart statistics arise in the context of modern heterogeneous data sources: "The term '*trusted* smart statistics' means, in particular, that the smart statistics can be relied on to provide reliable, robust, and accurate information, with the adjective 'trusted' implying that the decisions are based on sound data and information extraction" (Vichi and Hand, 2019). An extensive and enlightening discussion of trust in official statistics is given in Lehtonen (2019).

In other domains, trust in the statistical processing implicit in real-time and machine learning systems can, of course, be critical – think of autonomous vehicles for example (e.g. the Autonomous Chapter Event on safety and AI (Autonomous, 2020)). More generally, trust in artificial intelligence (AI) systems has been the focus of much discussion (e.g. EU, 2020). The increasing attention being paid to such areas is illustrated by the the recent conference on validating artificial intelligence systems (*ValidateAI*, 2019) and the recent call by UK Research and Innovation for grant applications for *Trustworthy Autonomous Systems Nodes* (UKRI, 2019) demonstrates.

As far as science is concerned, the European Federation of Academies of Sciences and Humanities (ALLEA) has set up a working group on truth, trust, and expertise, which has produced a series of publications (ALLEA, 2018a,b, 2019a,b) and run workshops and conferences to explore "current and past dynamics of public trust in expertise and the contested norms of what constitutes truth, facts, and evidence in scientific research and beyond."

Interest in trust has also been brought to the fore by the so-called "reproducibility crisis" in science, or at least within certain sciences. This arises from the fact that many apparent discoveries seem to vanish on re-examination. Hardly surprisingly, this leads to mistrust, not only in the conclusions themselves, but also in statistical analysis and even in the entire scientific method, as is illustrated by headlines such as "Is science broken?" (Rhodes, 2015).

Concepts of trust and trustworthiness are closely connected to ethical and governance issues, and an increasing amount of attention has also been paid to these in recent years (e.g. ASA, 2018; Hand, 2018b) with, for example, the UK government recently establishing a Center for Data Ethics and Innovation, the Royal Statistical Society launching a Special Interest Group in Data Ethics, and the UK's Office for National Statistics creating a Centre for Applied Data Ethics. Much of this is tied up with contracts, codes of conduct, encryption, monitoring, and so on; that is, with how people behave. Indeed, the UK Statistics Authority's Code of Practice explicitly defines trustworthiness as "confidence in the *people* and *organisations* that produce statistics and data" (my italics), and it is noteworthy that "methods" do not appear under trustworthiness, but under the quality pillar. Similarly, the first "Responsibility" given in the *Singapore Statement on Research Integrity* (Singapore, 2010) is "Integrity*:* Researchers should take responsibility for the trustworthiness of their research." Presumably this seeks to cover the researchers, the data, and the analysis tools, but in general guidelines, protocols, and codes of practice for researchers seem to place most emphasis on the trustworthiness of researchers themselves. For example, Responsibilities 9 and 11 of the Singapore Statement, to do with conflict of interest and reporting irresponsible research practices respectively, also discuss trust. Likewise, the first responsibility in the *Montreal Statement on Research Integrity* (Montreal, 2013) is "Integrity. Collaborating partners should take collective responsibility for the trustworthiness of the overall collaborative research and individual responsibility for the trustworthiness of their own contributions," and the second is "Trust. The behavior of each collaborating partner should be worthy of the trust of all other partners. Responsibility for establishing and maintaining this level of trust lies with all collaborating partners." Again the emphasis is on the people.

In short, we see that explorations of trust most commonly focus on trusting people and their behaviour: for further discussion see O'Neill (2002) or Hawley (2012) for general discussions, or Spiegelhalter (2017) within the context of statistics.

Emphasising people and their behaviour is all very well, but to have trust in a scientific (including statistical) conclusion we need not only trustworthy researchers but also trustworthy data and trustworthy data analysis.

Trust in the data is at risk from those who do not understand the limitations of their data or from those who deliberately set out to mislead. Aspects of trust in the data are discussed in Hand (2020). The bottom line is that one must have confidence in the way the data have been produced – in its *provenance* (Hand, 2018a). This includes matters of definition, incompleteness, measurement error, and all the other potential shortcomings. The familiar punchy adage *Garbage In, Garbage Out* summarises the consequences of not being certain about how the data have arisen and of not understanding their limitations. Or, to put it more positively, any conclusions from an analysis must be qualified by any uncertainties one may have about how the data have been generated. In this paper we take trustworthy data as a given.

Trust in the analysis, which is the concern of this paper, is likely to be at risk as a consequence of ignorance of the limitations of methods or of which methods are best suited to answer a particular question. This will also refer to more basic ignorance about software packages – like the recent example of failure to download 16,000 confirmed cases of Covid-19 because of lack of awareness of file size limitations in early versions of Excel (Vincent, 2020). In this paper we assume that the methods are understood, used properly, and interpreted correctly. Beyond that, however, the very nature of randomness means we need to know to what extent we can trust a conclusion or discovery, rather than it arising by accident as chance variation. The ability to evaluate this has been at the core of statistics since it became a formal discipline.

*We stress that this paper is not a review of the now very substantial literature on the use and misuse of hypothesis tests and p-values*. Rather our aim is to examine this work from the broader perspective of trust and trustworthiness, to see how the challenges to statistical testing and appeals for alternative approaches fit into this context. In particular, we examine the potential impact of proposals to ban the use of certain statistical tools on the discipline itself.

We argue that such a ban could have severely adverse consequences, both for individual analyses (the micro level) and for the discipline (the macro level). It threatens to undercut the basis for trust in statistical conclusions arising from the trustworthiness of the methods and it threatens to aggravate the already conflicted public view of the value of statistical analysis. We suggest that, rather than discouraging or prohibiting the use of the very tools which can lead to trust, we need to educate and regulate for the proper use of those tools.

We begin, in the next Section, by briefly exploring more closely what is meant by "trust" and "trustworthiness", especially the use of those terms in a statistical context. We note, in particular, the role of opening oneself to criticism as a way of demonstrating trustworthiness. Section 3 then goes on to discuss trust and trustworthiness in the statistical methods themselves before, in Section 4, describing the main underlying reasons for the criticisms of p-values and significance tests and examining the American Statistical Association comment on statistical significance and p-values (ASA, 2016) and more recent proposals to ban certain terms from statistical practice. Section 5 pulls the material together, arguing that statistical terms and methods can be useful when used properly, and making some recommendations to encourage this. This section points out that banning any potentially useful method – perhaps especially methods aimed at enabling trustworthiness – is shortsighted, unscientific, and Procrustean, that it damages the capability of science to advance, and, worse still, feeds into public mistrust of the discipline of statistics.

## 2. Trust and trustworthiness

The Oxford English Dictionary defines *trust* as a "firm belief in the reliability, truth, ability, or strength of someone or something". *Trustworthy* is defined as "able to be relied on as honest or truthful." So trust is our confidence (degree of belief) in the reliability, truth, etc of someone or something, while trustworthiness is the extent to which that person or thing merits trust. Trustworthiness is deserving of trust. This means we can seek evidence that someone or something is deserving of trust (is trustworthy), and hence whether we should place our trust in them.

Clearly these definitions nicely capture a key desirable attribute of our statistical methods. Above all, we require them to be "able to be relied upon as honest or truthful", that is, to be trustworthy, and it will follow from this that we would have a belief that the conclusions based on them can be trusted.

What might lead us to regard something as trustworthy, and hence to place our trust in it? The key will be that it has earned our trust or proven to be trustworthy based on our knowledge or experience of it. Note that such knowledge or experience might be transitive: someone we believe to be trustworthy might have asserted that something is in turn trustworthy. Knowledge or experience is what underpins the notion of *earning* our trust, and this has at its heart the concept of *evidence*. A method is trustworthy if we have examples of it satisfying the conditions above, and do not have examples of it failing to satisfy them.

In this vein, Onora O'Neill (O'Neill, 2012) says "I think it's perverse to think of trust as more basic than trustworthiness. To place and refuse trust intelligently we need first to judge others' trustworthiness, or their lack of trustworthiness, in specific matters. Only when we can do so, will we be in a position to place and refuse trust intelligently." Likewise Sekhon *et al* (2014) say

"we identify trustworthiness as a separate upstream construct with its own properties as the key antecedent of trust". Of course, both of these are writing about the trustworthiness of people, but the same applies to statistical methods. We can place or refuse trust in statistical conclusions intelligently when we have seen evidence of the trustworthiness or otherwise of the statistical methods leading to those conclusions. A method which experience has shown to often lead to incorrect conclusions would hardly be regarded as trustworthy, and we would have little confidence in any conclusions reached using such a method.

The distinction between trust and trustworthiness becomes important because, as O'Neill (2013) says: "Trust is the response, trustworthiness is what we have to judge."

If trustworthiness is what we have to judge, then we will look for evidence for it – or for its complement. Whyte and Crease (2010) point to one way of acquiring such evidence: "trust means deferring with comfort and confidence to others, about something beyond our knowledge or power, in ways that can potentially hurt us". O'Neill (2013) spells it out: "If you make yourself vulnerable to the other party, then that's very good evidence that you are trustworthy" and Spiegelhalter (2017) repeats it: "you also have to provide usable evidence that allows others to check whether you are trustworthy, which necessitates making yourself vulnerable to the other party." Put bluntly, this is saying "I grant you the power to penalise me if I do not do what I say I will do."

## 3. Trust in statistical methods

We saw in the previous section that, as far as people were concerned, making yourself vulnerable would provide evidence of your trustworthiness. This is translated into statistical inference by using a method which we know has a high probability of producing a consequence which is substantially discrepant from what we would expect were an assertion true, whenever the assertion is false. To take a familiar standard example, suppose I assert that some parameter of a system (say a mean or a correlation, for example) has a value different from zero. If this assertion is true, we might expect sample values of the statistic not to lie near zero (imagine making the sample size large and using a Wald statistic). Then the statistical analogy to vulnerability is provided by a test which has a high probability of producing a value close to zero if my assertion is false and the parameter is in fact zero. A test with this property, being vulnerable to the assertion being false, is trustworthy. It means that, were it to fail to produce a value close to zero, we would have no evidence for supposing the assertion false.

The "high probability" in the above discussion can be interpreted in at least two ways. The first, and probably most familiar, is closely related to the calibration of a method, defined by Reid and Cox (2015) as "the behaviour of a procedure under hypothetical repetition. That is, we study assessing uncertainty, as with other measuring devices, by assessing the performance of proposed methods under hypothetical repetition". They say: "The role of calibration seems essential: even if an empirical frequency-based view of probability is not used directly as a basis for inference; it is unacceptable if a procedure yielding regions of high probability in the sense of representing uncertain knowledge would, if used repeatedly, give systematically misleading conclusions." Certainly in such circumstances the method could not be regarded as trustworthy.

Alternatively, in Mayo's error-statistical severe-testing approach, the "high probability" does not depend upon notions of hypothetical repetition under identical conditions which are assumed to "rub off" on the particular instance, but rather the probability is taken as a measure of the strength of the test that has been passed (Mayo, 2018, p429). As Mayo puts it, "It's the sampling distribution of the given experiment that informs us of the capability of the method to have unearthed erroneous interpretations of the data" (Mayo, 2018, p429). If the method had a high capability of suggesting our hypothesis is wrong if it were wrong, but did not do so, then that is evidence that the hypothesis is right. "That's what it means to *view statistical inference as severe testing*. A claim is severely tested to the extent it has been subjected to and passes a test that probably would have found flaws, were they present" (Mayo, 2018, pxii). In summary, trust in a claim derives from the trustworthiness of the method used to establish it, and this trustworthiness lies in its high capacity for detecting spurious claims, its vulnerability to evidence to the contrary.

We can contrast "trustworthy" with "probable", since the word *probable* means something is likely to be true (or to happen), carrying no notion of the strength of justification for that assertion. This critical distinction allows Mayo to go on to say "The goal of *highly well tested* claims differs sufficiently from *highly probable* ones that you can have your cake and eat it too: retaining both for different contexts. Claims may be 'probable' (in whatever sense you choose) but terribly tested by these data. … The testing metaphor grows out of the idea that before we have evidence for a claim, it must have passed an analysis that could have found it flawed," (Mayo, 2018, pxii). Or, in our terms, the claim must be vulnerable to discrediting, should it be flawed. This is what allows Gelman and Shalizi (2013) to say that: "Implicit in the best Bayesian practice is a stance that has much in common with the error statistical approach of Mayo (1996), despite the latter's frequentist orientation."

The notion of trust via well-testedness is a familiar and widely-held one, although often not couched in terms of trust. Here are some examples. George Box comments "I believe that . . . sampling theory is needed for exploration and ultimate *criticism* of the entertained model in the light of the current data" (Box, 1980). It's the criticism, and the fact that a model has survived tough criticism, which ensures trust. Donald Rubin writes "frequency calculations are useful for making Bayesian statements scientific, scientific in the sense of capable of being shown wrong by empirical test" (Rubin, 1984). It's the capability of being shown wrong which is makes it vulnerable. Roderick Little says "the search for procedures with good frequentist properties provides some degree of protection against model misspecification" (Little, 2011). And Adrian de Groot says "Ceteris paribus, a theory or hypothesis is the more valuable as it risks more; its value will reach rockbottom if in the formulation no risk of refutation is incurred at all" (de Groot, 1969, p127).

Given that the trustworthiness of a method (and hence of the trust we put in its conclusions) resides in its capacity to flag as false those assertions which are false, two important tools in evaluating trustworthiness are the p-value and the significance test.

The p-value is defined as the probability of obtaining a statistic at least as "extreme" as that observed in the data, under the assumption that our assertion is false, and hence its complement true. The complement of the p-value, $1-p$, tells us the probability of obtaining a test statistic value less extreme than the observed value if our assertion is false. Taking the example above, the assertion was that the parameter had a value different from zero. The complement of the p-value tells us what would be the probability of obtaining a statistic with absolute value smaller than that actually observed if that non-zero assertion is false, so that the parameter actually has value zero. If this probability is large, and hence the p-value small, the test had a high capacity to flag our assertion as false – but it did not do so. We have evidence that the assertion is true. Of course, the validity of the p-value also depends on other model assumptions holding, so single rejections cannot be definitive – we are, after all, discussing the complexity of the real world. Ronald Fisher wrote that "we may say that a phenomenon is experimentally demonstrable when we know how to conduct an experiment which will rarely fail to give us a statistically significant result" (Fisher 1935, p. 16). The more such experiments are conducted, with other model assumptions varying, the more confidence we can have in our assertion.

A test which has a small p-value has certainly demonstrated the requisite vulnerability requirement: if my assertion (non-zero parameter) is false (and hence its complement – parameter equals zero – is true) then the test had a large probability (1 minus the p-value) of producing a value closer to zero than that actually observed.

We need to quantify what we mean by a "small" p-value. We could choose some threshold $t$ so that an observed p-value smaller than $t$ would imply that the probability of obtaining less extreme results was greater than $1-t$ were our assertion false. If we observe a p-value smaller than $t$ we say that the result is *significant*. The choice of this threshold will depend on the context and the confidence one wishes to have in the conclusion. Even if a threshold is not chosen explicitly, it will often be chosen implicitly: we will want to know if the result constitutes strong evidence against the hypothesis, and we will have some idea of what strength is sufficient to be regarded as compelling. The use of an explicit threshold is simply a convenient way of informally defining what the researcher considers to constitute a sufficient degree of trustworthiness for the purposes of the particular study.

Note that the observed p-value and whether or not it is lower than some chosen threshold serve two rather different purposes. The former is useful for gauging the strength, and hence degree of trustworthiness, of the test that has been passed, and as a consequence also the degree of trust that might be placed in a conclusion. In contrast, the latter tells us if a trustworthiness threshold has been reached, so that it is useful for guiding possible actions: does our confidence in the results reach a level adequate for our purpose?

**4. Undercutting trust**

The outline of p-values and their role in the trust we should put in statistical conclusions given in earlier sections seems both straightforward and logically sound, so it is perhaps surprising that there has been a recent outpouring of publications criticising p-values and in particular their use with a threshold level. Such criticisms have been made for decades, generating a huge literature but they have gained new force with the recent growth in interest in the reproducibility crisis mentioned in the introduction. Given the size of the literature, the reader must forgive me for not attempting an extensive discussion. Not only would that unbalance this paper, but it may be impossible: McShane *et al* (2019) say, "the breadth of the literature on this topic across time and fields makes a complete review intractable". Instead I have focussed on just a few matters which are particularly pertinent to the trust/trustworthiness perspective described in this paper. None of these are new, and a reader familiar with this debate could skip this section. For a broader discussion see Mayo and Hand (2021).

In an attempt to cut through the thicket of the debate, the American Statistical Association published a "statement on statistical significance and p-values" (ASA, 2016). The statement, which was supplemented by an extensive collection of online comments, said "[the p-value] is commonly misused and misinterpreted" and then went on to say "This has led to some scientific journals discouraging the use of *p*-values, and some scientists and statisticians recommending their abandonment, ….". The first of these statements (about the misuse and misinterpretation) is undoubtedly true. Greenland *et al* (2016) contains a good review of the mistaken understanding and uses of p-values and McShane and Gal (2017) demonstrates its extent. Of course, misuse and misinterpretation do not necessarily have adverse consequences if they are near enough to what is needed, and we should not mistake a simplified shorthand way of describing something as necessarily implying misunderstanding (c.f. we speak of the sun "rising"). However, the reaction by some journals, scientists, and statisticians described in the second statement is curious and unscientific. Should the fact that a tool can be misused mean it should be abandoned, regardless of its merits, strengths, value, and usefulness? We do not say the same for cars, aircraft, knives, or ropes, all of which can be misused. Instead we educate and regulate for the proper use. The ASA statement continues "In this context, the American Statistical Association (ASA) believes that the scientific community could benefit from a formal statement clarifying several widely agreed upon principles underlying the proper use and interpretation of the *p*-value." This is, of course, entirely beneficial and to be applauded.

Despite its limited aims, the ASA statement represented a useful contribution in pointing out the role of p-values and also drawing attention to some of the misunderstandings. It is noteworthy, however, that the statement did not include the word which is at the heart of why p-values are so useful (at least, when used and interpreted properly): that is *trust*, in the statistical conclusions.

The third recommendation in the ASA Statement is "Scientific conclusions and business or policy decisions should not be based *only* on whether a *p*-value passes a specific threshold." (my italics) That is, in other words, "should not be based *only* on whether the result was statistically significant". Presumably no-one would disagree with the assertion that such conclusions should not be based *only* on a single characteristic of the data. However, "should not" is of course not the same as "will not" or "is not", and unfortunately exceeding a threshold is used all too often as an indicator of discovery or even "truth". There are two aspects to this mistaken usage. One is the familiar one of failure to understand the role and limitations of p-values and significance tests – the misuse and misunderstanding mentioned throughout this paper. The other is a failure to appreciate the distinction between scientific questions and their representation as statistical questions. We discuss this in the concluding section.

Asserting principles (and generally accepted principles at that) is one thing. But it is another to seek to impose bans on the use of tools for facilitating scientific discovery. A successor paper to the ASA statement (Wasserstein *et al*, 2019), went on to do this. Focusing on the practice of comparing p-values with a pre-specified threshold, they say "it is time to stop using the term 'statistically significant' entirely". The nuanced recommendation in the earlier 2016 statement, that such tools might have a place, but are probably not the *only* relevant information, has been discarded. Wasserstein *et al* do not recommend stopping using p-values: "we are not recommending that the calculation and use of continuous p-values be discontinued", though they do say "results should not be trichotomized, or indeed categorized into any number of groups, based on arbitrary p-value thresholds". Again, singling out the word *arbitrary*, it is difficult to disagree with this. Arbitrariness surely has no place here, and certainly risks damaging trustworthiness in what is being done. But these authors mean something more, as we explore below.

Incidentally, regardless of the merits of the p-value/significance debate, one might argue that prohibiting the use of any (ethically acceptable) scientific tool represents an unreasonable constraint on the discovery process if that tool has proper and valid uses, even if it is often misused. Rather than removing a valuable and effective tool, would it not be far better to ensure that it is used properly? The old adage about babies and bathwater comes to mind.

In a special issue of *The American Statistician* aimed at helping researchers undertake effective and valid statistical analysis, Wasserstein *et al* (2019) is accompanied by 43 other papers. These papers represent a diversity of views, some suggesting alternatives to significance testing and p-values, others suggesting entirely different approaches, and yet others supporting them. This means that these 43 papers are sometimes contradictory (Wasserstein *et al*, 2019, say "At times in this editorial and the papers you'll hear deep dissonance"). Debate, of course, lies at the heart of science, so it is good that this material has appeared. John Milton put it nicely in 1644: "Where there is much desire to learn, there of necessity will be much arguing, much writing, many opinions" (Milton, 1644). At the least, this material has the potential to demonstrate that statistics is a broad and dynamic discipline, where problems can be approached in more than one way, and where different tools have different and sometimes complementary properties. On the other hand, if the different approaches are described in an antagonistic way (and there is plenty of that in this literature) it is bad for the discipline and its reputation. It could have the consequence of leading the less statistically informed observers of the discussion to mistrust statistical methods in general, a point we return to below.

Even if Wasserstein *et al* (2019) do not recommend banning p-values per se, other authors have suggested that this might be a direction to proceed. For example, Gelman and Carlin (2017) suggest "Another direction for reform is to preserve the idea of hypothesis testing but to abandon tail-area probabilities (*p*-values) and instead summarize inference by the posterior probabilities of the null and alternative models …" (but then they point out the difficulties with this strategy).

More seriously, in a bizarre move to undercut trust in the material it publishes, the journal of *Basic and Applied Social Psychology* decided to ban "null hypothesis significance testing" (Trafimow and Marks, 2015). In answer to two of the questions about how the ban would be implemented, these authors gave the following replies:

*Question 1.* Will manuscripts with p-values be desk rejected automatically?

*Answer to Question 1.* No. If manuscripts pass the preliminary inspection, they will be sent out for review. But prior to publication, authors will have to remove all vestiges of the NHSTP (p-values, t-values, F-values, statements about ''significant'' differences or lack thereof, and so on).

*Question 3.* Are any inferential statistical procedures required?

*Answer to Question 3.* No, because the state of the art remains uncertain.

I'm sure it was not their intention, but the answer to Question 1 seems to be inviting authors to hide the steps they went through to reach a conclusion. In my refereeing experience, I have often had to ask authors to spell out just how they reached a conclusion, or give more information so that I could properly assess a claim. I have certainly never requested that an author should conceal the procedures they undertook! The proposed practice seems both ethically dubious and almost diametrically opposed to promoting trustworthiness.

Likewise, the answer to Question 3 seems equally curious. If we were to wait until we had certainty, either in methodology or in conclusions, we would wait forever. It is fundamental to science that its conclusions are contingent upon more data becoming available which might cast doubt on current understanding. This is as true of statistical science as it is of psychological science.

This curious policy appears to be based on an error of fact: Trafimow states that "The null hypothesis significance testing procedure has been shown to be logically invalid" (Trafimow, 2014, and repeated in Trafimov and Marks, 2015). He is right that such testing does "provide little information about the actual likelihood of either the null or experimental hypothesis", but it is not intended to do that, so his criticisms are misdirected at a straw man rather than at the reality. Admittedly he is not alone in making this mistake, but journal editors have a particular responsibility to understand the tools that researchers might use.

Strangely, Trafimow also says "Experiments should include sufficient participants so as to instill some confidence in the stability of obtained effect sizes". Isn't "instilling confidence" despite random variablility the essence of statistical inference? And surely this is exactly what p-values do, by relating the observed effect size and its variation to a hypothesised effect size. Fisher (1959, p76) wrote: "[tests of significance] are constantly in use to distinguish real effects of importance to a research programme from such apparent effects as might have appeared in consequence of errors of random sampling, or of uncontrolled variability, of any sort, in the physical or biological material under examination" and Greenland *et al* (2016) say "[Statistical tests] were originally intended to account for random variability as a source of error, *thereby sounding a note of caution against overinterpretation of observed associations as true effects or as stronger evidence against null hypotheses than was warranted.*" (my italics). That is, statistical tests aim to "instill some confidence" – or since, as we said in Section 2, trust is our confidence in the reliability, truth, etc of something, and p-values enable trust.

Fricker *et al* (2019) examined the impact of the ban on the publications appearing in *Basic and Applied Social Psychology*. They say "failing to first skeptically assess whether an observed effect could be consistent with random variation may result in an excessive number of false positives in research results". The consequence may be mistrust of the results described in the journal. And they say "Indeed, the scientific method demands skepticism of any observed results, where one should first want to rule out the simplest explanation that an observed result is consistent with random variation in the data before then seeking to find another explanation." We are back to the notion that the procedure must be *vulnerable* to validation if it is to be trustworthy.

Fricker *et al* do not give a quantitative assessment but Academic Accelerator (2020) notes that "The Journal Impact 2019-2020 of *Basic and Applied Social Psychology* is 1.290, which is just updated in 2020. Compared with historical Journal Impact data, the Metric 2019 of Basic and Applied Social Psychology dropped by 66.58%". Of course, one might legitimately doubt the value of impact factors as a measure of the merit of a journal or its content, not least because they can be manipulated. Likewise submission rates to a journal implementing such a ban might be expected to go down because authors recognise the fact that limited trust can reside in conclusions published in its papers, or to go up because the significance test gatekeeper has been abandoned. Either way it damages trust in science.

Amrhein *et al* (2019) draw attention to a bias in the scientific literature arising from the pressure (from various directions) to privilege significant results. This is doubtless one of the underlying contributors to the reproducibility crisis mentioned in Section 1 (see also Hand, 2020). They discuss various remedies which have been proposed to alleviate this situation, including pre-registration of studies and a commitment to publish all results of analyses, but point out that even these remedies can be influenced. This leads them to the suggestion that the solution is a ban on the use of statistical significance. But surely their argument is based on a false and oversimplified premise. It is not the pressure to privilege significant results per se which leads to the bias, but what underlies this: a (natural?) desire for recognition and reputation by means of theories and discoveries which are widely accepted by the scientific community. Significance tests are but a flag which indicates that such a discovery might have been made. And some sort of flag or indicator or gatekeeper is necessary.

Addressing the question "what will retiring statistical significance look like?" Amrhein *et al* (2019) give a list, preceded by "we hope that". Many of these aspirations would surely be held by most statisticians and so seem irrelevant to a proposal to retire significance testing (which is the title of the paper). For example, their list includes: emphasise estimates and the uncertainty in them; give more details about methods; report p-values with sensible precision; spend less time with statistical software and more time thinking. But, given the extraordinarily widespread use of significance testing, the proposal has huge practical implications. That being the case, and given that the proposal is to sacrifice some (not perfect, but some) protection against drawing conclusions based on chance variability, do we not require exceptionally strong evidence that the policy will have a beneficial effect. Is Amrhein *et al*'s "we hope that" sufficient? It is somewhat ironic that a discussion about the nature of the validity of empirical evidence should itself be replete with proposals but lack evidence. Such a situation can hardly be conducive to high level trust in the discipline. The proposals seem readily accessible to experimental investigation. So I suggest that any journal editor contemplating a ban on significance testing (or any other statistical tool for that matter) should first consider whether the strategy could be properly evaluated in a controlled experiment. I believe we missed an opportunity with those journals which have already implemented such a ban. At best we can now collect observational data.

The need for concrete evidence is further illustrated by the admission that, although Amrhein *et al*'s call to retire statistical significance might eliminate some bad practices (they say "will", but, again as far as I can see, without evidence), "it could well introduce new ones". To make an informed judgement about whether the risks of the action they propose are justified, we need to explore the potential downsides. To do otherwise risks unnecessarily damaging trust in our discipline in unknown and unquantified ways. In particular, it is possible that the freedom to give researchers carte blanche to characterise any magnitude of p-value as providing evidence for some position can only aggravate the bias Amrhein *et al* refer to. As Ioannidis (2019) puts it, "Absent prespecified rules, most research designs and analyses have enough leeway to manipulate the data and hack the results to claim important signals." In short, again, abandoning significance levels threatens trust.

McShane *et al* (2019) also discuss problems with the use of "null hypothesis significance testing", including noisy estimates, point null hypotheses, bias in reporting, and selection bias from noisy estimates. Surely, problems of excessive variation arising from small samples are not a problem of significance testing. They are a problem of experimental design or of failing to take the potential variation into account when drawing conclusions. Again, significance tests serve as a gatekeeper (again albeit not perfect) against this, and there is a real danger that, without the formality of such tests and their clear thresholds, greater selection bias will occur. As to the majority of applications adopting a sharp point null hypothesis, if a sharp point null hypothesis is not appropriate for a particular study then a sharp point null hypothesis should not be used. But this is not the fault of the method, but of how it is used – and probably the person misusing it.

Other problems mentioned by McShane *et al* (2019) (overlapping with objections raised by others, as one might expect), with my comments below them, include:

- that categorisation of results into significant and not significant encourages researchers to interpret evidence dichotomously rather than continuously.

  Presumably this is analogous to the way that using the arithmetic mean encourages researchers to ignore extreme values and skewness. It is the responsibility of the researcher to understand the tools they are using, and their properties and limitations. It is the responsibility of educators to ensure that they do. It is the responsibility of journal editors to ensure that the material they publish use methods correctly, insofar as they can tell.

- arbitrariness of the conventional 0.05 threshold.

  Yes, the 0.05 threshold is arbitrary, and it should not be used without a carefully spelt out justification. However, as noted above, dropping explicit thresholds altogether is equivalent to allowing researchers to decide arbitrarily, and perhaps in an obscure way, what they regard as worth noting. Requiring explicit threshold specification *and justification*, not off-the-shelf blanket 0.05 or 0.01, forces thought. Fisher (1959 p42) was clear about the proper use: "No scientific worker has [in the light of 60 years of experience, perhaps we should say "should have"] a fixed level of significance at which, from year to year, and in all circumstances, he (sic) rejects hypotheses; he rather gives his mind to each particular case in the light of his evidence and his ideas". Use of any tool without justification hardly promotes trustworthiness.

- researchers take rejection of a sharp null hypothesis as positive evidence of some particular alternative.

  A common misunderstanding – but surely a fault of the users (and, by implication, educators) and one which is relatively easy to remedy.

- failure to take a holistic view, including all relevant evidence, not merely a test result.

  Presumably this is also primarily a failure in education. Researchers should be taught that the scientific enterprise is not a mechanical matter, with numbers being thrown into a statistical machine which outputs scientific conclusions. They should recognise that multiple sources of evidence are generally relevant.

- confusion of statistical and practical significance.

  This is hardly the fault of significance tests themselves, though it might form the basis of an argument for using a different word for "statistical significance".

- misinterpretations of the p-value (such as taking it to mean the probability that the null hypothesis is true).

  Again, failure to understand the basic concepts and tools being used can hardly be attributed to those tools. If I took control of a passenger jet and accidentally crashed it, you would probably and rightly attribute the crash to my lack of understanding of how to fly it, rather than any shortcomings of the aircraft itself.

I agree with McShane *et al* (2019) when they say in Section 4.1, "Statistics is hard, especially when effects are small and variable and measurements are noisy." Of course, those are also amongst the situations which most need statistics. I likewise agree that "a formulaic approach to statistics is a principal cause of the current replication crisis". Formulaic, "turn the handle", approaches have no place in statistics. But that criticism is orthogonal to any criticisms of p-values and significance, apart from the fact that mistaken interpretations and use of p-values are an example of formulaic uses. Other statistical tools are also misused because of convention and simple lack of thought (see Hand, 1994, for examples). But I disagree with McShane *et al* when they say that each of the proposals for tackling the misunderstandings and misuse of significance (and retaining the use of the tool) "is a form of statistical alchemy that falsely promises to transmute randomness into certainty" ("uncertainty laundering", Gelman, 2016). If a researcher believes a significant p-value indicates certainty then they have misunderstood significance testing. Moreover, when they go on to say "There are no quick fixes" I wonder how that sits with their call to "abandon statistical significance" (the title of their paper), which seems to me to be the ultimate in quick fixes.

Trustworthy inference, valid inference and statistical tests, also require that the source and quality of the data are known, that results must not be cherry-picked, that multiple testing must be accounted for, that assumptions are justifiable, that the sample was properly drawn, that researcher degrees of freedom are taken into account, and so on. To the extent that poor scientific practices occur, the methods are untrustworthy and the results should not be trusted. But criticisms of the way a tool is used should be laid at the feet of those wielding the tool, not at the tool itself. As Benjamini (2016) put it "it's not the p-values' fault".

In summary, criticisms of p-values and significance tests seem to hinge upon widespread misunderstanding of their proper use and interpretation, disparaging them for not doing something they were never intended to do, their failure to provide information when they were not intended to provide, and in general conflating shortcomings arising from poor scientific practice with shortcomings of statistical procedures. While banning the use of significance testing would certainly mean such tests could not be held responsible for mistakes arising from these sources, neither would the tests then be able to provide the information they can provide when used properly. Worse still, the potential downside, in terms of the bad practices of strategies which might replace them, is unquantified. At the very least it would cast a pall of mistrust over our discipline.

## 5. Discussion

Trust in statistical tools and their use has often been lacking amongst the general public, and the discipline has always suffered from misunderstandings and misplaced criticism. Aphorisms such as "There are lies, damned lies, and statistics", "You can prove anything with statistics", "There are two kinds of statistics: the kind you look up and the kind you make up", and so on are commonplace. The situation has not been helped by the rise of concepts such as the "false facts" and "fake news" mentioned above.

Such trust has also long been lacking in some sectors of the scientific community, though there it is complicated by the existence of several different schools of statistics with fundamentally different ideas of how inference should be carried out, and indeed of what "probability" means.

Perhaps the root of the problem lies in the fact that researchers would often (and perhaps understandably) like an automatic way to make inferences: they would like a handle to turn to automate scientific discovery and statistical analysis based on their data. The challenge in achieving this is that to answer a scientific question a mapping must be established from that question to a statistical question. This is typically a difficult process, with much scope for ambiguity and confusion (see the examples in Hand, 1994, 1997, 2012). Simplifications are necessarily involved – a point that Trafimow (2019) makes, but then uses in an ultimately reductive sense as a sign that the entire p-value enterprise is doomed because it cannot allow for *all* the assumptions involved in the mapping, as no formal representation can possibly do. It is in the nature of scientific models that they simplify, and it is up to the scientist to attempt to ensure that irrelevant discrepancies are sufficiently small so as not to impact the conclusions. The history of science is largely one of carefully controlling for distorting factors.

This opportunity for ambiguity and confusion is illustrated for econometrics papers by a searing indictment of their trustworthiness from Aris Spanos (Spanos, 2010), who suggests that "the primary potential sources of error contributing to the untrustworthiness of evidence" include inaccurate data, inappropriate measurements, inadequate inferential analysis, and inadequacies in the scientific formulations of the questions, and then goes on to identify the emphasis of theory over data as "the single most important contributor to the untrustworthiness of empirical evidence in economics".

How to go about this mapping from the real world to the statistical model is usually glossed over in teaching (statistics), not least because it is intrinsically context-dependent, so that generalisations are difficult. Even in statistics texts aimed at researchers in other disciplines, very little, if anything, is said about the distinction between the scientific and the statistical problem. It has to be said that we statisticians might be partly responsible for this. Statistics has a history of having been taught within mathematics departments of universities, so that the instruction typically begins with a given data set and the mathematics (or algorithms) of the methods. This despite the fundamentally opposed aims of the two disciplines (in caricature: the aim of mathematics is to deduce the observable consequences in an artificial world described by a set of axioms, and the aim of statistics is to discover the nature of the world from its consequences (the data)). But the real world is not a mathematical world defined by a bunch of axioms. It is far more complex – as Sir David Cox put it: "The idea that complex physical, biological or sociological systems can be exactly described by a few formulae is patently absurd" (Cox, 1995). The real world is beset by an unlimited number of interacting variables, of which only a few will be measured, and measured with error at that, while definitions might be ambiguous or differ between studies, populations might differ or be uncertainly specified, nonstationarity and correlation can contaminate raw data, the data might be incomplete or drawn in under-specified ways, and so on and on.

Moreover, as Sir Austin Bradford Hill wrote (Hill, 1965) that "All scientific work is incomplete – whether it be observational or experimental. All scientific work is liable to be upset or modified by advancing knowledge." That, indeed, is the defining feature of science, and what distinguishes science from religion (where "truth" is given) and pure mathematics (where the axioms define the "world" being studied). In science new evidence can upset existing theories, requiring them to be replaced or elaborated. The crucial test of a new theory is that it can explain both new evidence, that is new data, as well as past data. To the extent that it can do so, it is a better theory. In fact "test" is the key word here, a comparison of the data with the theory: the vulnerability to refutation by a test being what promotes trust in a conclusion.

This fundamentally contingent nature of science means that "best explanations" should be expected to change. New evidence – new data in our context – will sometimes mean that a previously generally accepted theory no longer provides an adequate explanation for observed phenomena and must be changed. This so-called "flip-flopping" of science is sometimes uncomfortable to lay people (as well as being uncomfortable to the scientist whose favoured theory has just been found wanting!). And, of course, the change should not be expected to be smooth and monotonic. Given the complexity of the real world, and the inevitably inaccuracies and shortcomings of our measurements, precise matches between data and the predictions of a theory are usually impossible. Instead we must rely on measures of compatibility. And this is (one place) where statistics comes in. In particular, statistical tests explore the extent to which discrepancies could be explained by chance variation. A discrepancy of such a magnitude that it was highly unlikely to be due to chance would cast doubt on a theory. The capacity of a test to produce such discrepant results when the theory is false is what makes it trustworthy, and the trustworthiness of the test means we can have confidence in its conclusions.

P-values and significance tests – *when properly used and interpreted* – are prime examples of intrinsically trustworthy methods. P-values instantiate the vulnerability of an assertion to being false. However, we stress that any method, no matter how intrinsically trustworthy it is in itself, is only trustworthy to the extent that it is used properly. Unfortunately, the extent of misuse and misunderstanding of p-values is such that many have doubts about the pragmatic trustworthiness of these tools, believing the point has been reached at which they should be abandoned, and replaced by other tools. There is no doubt that the misuse, coupled with the sociological drivers of science (e.g. the desire to produce results which gain recognition and approbation) has aggravated the extent of flip-flopping. We can tackle statistical misunderstandings, but the sociological drivers require a broader strategy, beyond this paper.

Unfortunately, as one might expect, none of the (huge variety of different) proposed alternatives do the same job as significance tests or p-values, and all of them have their own shortcomings. (We note parenthetically that since they provide different information they might well be useful in addition to p-values and significance tests.) However, abandoning something which uniquely sheds light on a particular aspect of the data and its relationship to a hypothesis (their gatekeeper roles in the face of sampling uncertainty) is treating the symptom, not the cause.

Given that p-values are intrinsically trustworthy if used properly, and given that they do a key job, surely a better solution would be to ensure that they *are* used properly. As we said above, we do not discard knives because they can be misused – and then try to cut our food with a fork or spoon. The problem is nicely encapsulated by Cohen (1994) when he claims "that NHST (null hypothesis significance testing) has not only failed to support the advance of psychology as a science but has also seriously impeded it." At first glance he appears to be attributing agency to the abstract mathematics of the NHST, which is nonsensical. What he really means is "… *the (mis)use of* NHST has not only failed …". The fact that, in this case, the misuse might have been egregious and extensive, is not an argument for abandoning a valuable tool, but an argument for properly training those who would use it. Frick (1996) says (on the subject of significance testing) "Its continued use is typically attributed to experimenters' ignorance, misunderstanding, laziness, or adherence to tradition", rather than any intrinsic merit of the concept. Perhaps, in the light of its continued evident value and use for some problems, *even if it is not perfect for all problems, as no method can be*, rather than simply condemning it, we would be better off characterising its value and identifying those circumstances, conditions, and situations in which it is an effective and useful tool, and then ensuring that researchers understood the situations in which it could be used and used it properly.

Given the extent of the literature discussing the pros and cons of significance tests and p-values, it is unlikely that I can come up with any new proposals for how to improve things, but I summarise my recommendations here:

- education of researchers in the proper use and understanding of the tools they would use is critical. This is as true of statistics as it is of chemistry, physics, or any other scientific discipline.

- journal editors have a responsibility to ensure that results described in their pages have been derived using proper instruments. They should require authors to say how many tests were conducted altogether, what data were discarded, what assumptions were checked, what statistical tests were considered and why, and so on. Editors have not shied from requiring authors to present their material in rigidly constrained ways – think of the standard introduction/methods/conclusion structure of papers required in some disciplines (Biology, 2003), or the 428 page Publication Manual of the American Psychological Association (APA, 2020a), so I see no reason why more should not be required in terms of these statistical and indeed scientific fundamentals. With the advent of the web, limited page-space is no longer an excuse for not giving full details of experimental work.

- in particular, in the current context, this means editors should require authors to justify their choice of threshold when conducting significance tests. "I chose 5% because everyone else used it in the past" would not be an adequate reason. This does not seem too harsh a requirement given the very stringent constaints made on other aspects of style and presentation made by some journals, such as those of the APA noted above.

- collect evidence on the effect of banning significance testing, and on the use of "alternatives" proposed. Explore whether it makes the situation better, or worse.

- a referee raised the interesting question of whether the extent of misuse is so great that it is too late to fix the situation through education (of researchers and editors) and through more rigorous publication practices. Lacking empirical evidence one way or the other, all I can say is that I do not believe so. Although there have been extensive calls for action from within the statistical community, this community is tiny compared with those which use p-values and significance tests. Perhaps we statisticians should produce a kitemark to which journals could aspire, indicating that they have reached a level of quality in terms of the statistical understanding and explanation they require of their authors.

I had hoped that the rebranding of statistics as the major component of data science would enable us to escape from misunderstandings such as those quoted at the start of this section. However, recent proposals for banning certain tools from within the community risk appearing to imply that the statistical community accept that those quotations represent the truth about the discipline, rather than that misuse of its tools is what underlies them. If such a distorted perspective on statistics becomes widely accepted it would represent the most dramatic example of a scientific discipline shooting itself in the foot. The consequences could be serious, in terms of damage beyond the discipline of statistics, to science, to public policy, to industry, to medicine, and everywhere that statistical tools are used – which is just about everywhere.

## Acknowledgements

I am greatly indebted to two anonymous referees, whose careful critique of earlier versions of this paper led me to appreciate its multiple shortcomings and drastically revise it. Any remaining shortcomings are, of course, entirely my fault.

## References

Academic Accelerator (2020) https://academic-accelerator.com/Impact-Factor-IF/Basic-and-Applied-Social-Psychology  Accessed 15 October 2020.

ALLEA (2018a) *Science in Times of Challenged Trust and Expertise*. https://www.allea.org/wp-content/uploads/2019/01/ALLEA-ConferenceProceedingsDigital.pdf  Accessed 16 October 2020.

ALLEA (2018b) *Loss of Trust? Loss of Trustworthiness? Truth and Expertise Today.* http://www.allea.org/wp-content/uploads/2018/05/ALLEA_Discussion_Paper_1_Truth_and_Expertise_Today-digital.pdf Accessed 16 October 2020.

ALLEA (2019a) *Trust Within Science: Dynamics and Norms of Knowledge Production*. https://www.allea.org/wp-content/uploads/2019/01/ALLEA_Discussion_Paper_2.pdf Accessed 16 October 2020.

ALLEA (2019b) *Trust in Science and Changing Landscapes of Communication*. https://www.allea.org/wp-content/uploads/2019/01/ALLEA_Trust_in_Science_and_Changing_Landscapes_of_Communication-1.pdf Accessed 12 October 2020.

Amrhein V., Greenland S., and McShane B. and over 800 other signatories (2019) Retire statistical significance. *Nature*, **567**, 305-307.

APA (2020a) *Publication Manual of the American Psychological Association*. American Psychological Assocation

ASA (2016) ASA statement on statistical significance and p-values. *The American Statistician*, **70**, 131-133.

ASA (2018) *Ethical Guidelines for Statistical Practice*. https://www.amstat.org/asa/files/pdfs/EthicalGuidelines.pdf

Autonomous (2020) https://www.youtube.com/watch?v=Z4jnqJtNOuo&t=13m50s Accessed 13 August 2020.

Benjamini Y. (2016) "It's not the p-values' fault". *The American Statistician, Online Supplement*. https://amstat.tandfonline.com/doi/suppl/10.1080/00031305.2016.1154108/suppl_file/utas_a_1154108_sm5354.pdf

Biology (2003) http://biology.kenyon.edu/Bio_InfoLit/how/page2.html Accessed 16 October 2020.

Box G.E.P. (1980) Sampling and Bayes' inference in scientific modelling and robustness. *Journal of the Royal Statistical Society, Series A*, **143**, 383-430.

Cohen J. (1994) The Earth Is round ($p$ < .05). *American Psychologist*, **49**, 997–1003.

Cox D.R. (1995) Comment on "Model uncertainty, data mining, and statistical inference". *Journal of the Royal Statistical Society*, **158**, 455-456.

de Groot, A. D. (1969). *Methodology: Foundations of Inference and Research in the Behavioural Sciences*. The Hague: Mouton & Co.

EU (2020) *White Paper: On Artificial Intelligence – A European Approach to Excellence and Trust*. https://ec.europa.eu/info/sites/info/files/commission-white-paper-artificial-intelligence-feb2020_en.pdf  Accessed 12 October 2020.

Fisher R.A. (1935) *The Design of Experiments.* Oliver and Boyd, London.

Fisher RA (1959) *Statistical methods and scientific inference, 2ⁿᵈ ed*. Oliver and Boyd, Edinburgh

Frick R.W. (1996) The appropriate use of null hypothesis testing. *Psychological Methods*, **1**, 379-390.

Fricker R.D.Jr., Burke K., Han X., and Woodall W.H. (2019) Assessing the statistical analyses used in *Basic and Applied Social Psychology* after their p-value ban. *The American Statistician*, **73: sup 1**, 374-384.

Gelman A. (2016) The problems with p-valus are not just with p-values. *The American Statistician*, **70**, 10 (supplemental material to the ASA statement on *p*-values and statistical significance).

Gelman A. and Shalizi C.R. (2013) Philosophy and the practice of Bayesian statistics. *British Journal of Mathematical and Statistical Psychology*, **66**, 8-38.

Greenland S., Senn S.J., Rothman K.J., Carlin J.B., Poole C., Goodman S.N., and Altman D. (2016) Statistical tests, P-values, confidence intervals, and power: a guide to misinterpretations. *The American Statistician, Online Supplement,* https://link.springer.com/article/10.1007/s10654-016-0149-3  Accessed 16 October 2020.

Hand D.J. (1994) Deconstructing statistical questions (with discussion), *Journal of the Royal Statistical Society*, *Series A*, **157**, 317–356.

Hand D.J. (1997) Scientific and statistical hypotheses:  bridging the gap.  *Understanding Social Research: perspectives on methodology and practice,* ed G.McKenzie, J.Powell, and R.Usher, Falmer Press, 124-136.

Hand D.J. (2012) Assessing the performance of classification methods. *International Statistical Review*, **80**, 400-414.

Hand D.J. (2018a) Who told you that?: data provenance, false facts, and separating the liars from the truth-tellers. *Significance*, August 2018, 8-9.

Hand D.J. (2018b) Aspects of data ethics in a changing world: where are we now? *Big Data*, **6**, 176-190. No.3, 176-191.

Hand D.J. (2020) *Dark Data: Why What You Don't Know Matters*. Princeton University Press.

Hawley K (2012) *Trust: A Very Short Introduction*. Oxford University Press, Oxford.

Hill A.B. (1965) The environment and disease: association or causation? *Proceedings of the Royal Society of Medicine*, 58 (1965), 295-300.

Ioannidis J.P.A. (2019) The importance of predefined rules and prespecified statistical analyses. *Journal of the American Medical Association*, **321(21)**, 2067-2068

Lehtonen M. (2019) The multiple faces of trust in statistics and indicators: a case for healthy mistrust and distrust. *Statistical Journal of the IAOS*, **35**, 539-548.

Little R. (2011) Calibrated Bayes, for statistics in general and missing data in particular. *Statistical Science*, **26**, 162-174.

Mayo D.G. (2018) *Statistical Inference as Severe Testing: How To Get Beyond The Statistics Wars*. Cambridge University Press, Cambridge.

Mayo DG and Hand DJ (2021) *Statistical Significance Tests: Practicing Damaging Science or Damaging Scientific Practice?* Submitted.

McShane B.B. and Gal D. (2017) Statistical significance and dichotomization of evidence. *Journal of the American Statistical Association*, **112**, 885-908.

McShane B.B., Gal D., Gelman A., Robert C., and Tackett J.L. (2019) Abandon statistical significance. *The American Statistician*, **73**, 235-245.

Milton J. (1644) *Areopagitica, A Speech of Mr. John Milton for the Liberty of Unlicenc'd Printing to the Parliament of England* (1 ed.). London. https://books.google.co.uk/books?id=nejQAAAAMAAJ&dq=areopagitica&pg=PP13&redir_esc=y#v=onepage&q=areopagitica&f=false Accessed 15 October 2020.

Montreal (2013) https://wcrif.org/documents/354-montreal-statement-english/file Accessed 12 Octoerb 2020.

O'Neill, O. (2002) *A Question of Trust: the BBC Reith Lectures 2002*. Cambridge: Cambridge University Press.

O'Neill O (2012) A point of view: which comes first – trust or trustworthiness? *BBC Point of View,* December 2012.

O'Neill O. (2013) https://www.ted.com/talks/onora_o_neill_what_we_don_t_understand_about_trust#t-356579 Accessed 16 October 2020.

PARIS21 (2020) https://trustinitiative2020.paris21.org/  Accessed 16 October 2020.

Reid N. and Cox D.R. (2015) On some principles of statistical inference. *International Statistical Review*, **83**, 293-308.

Rhodes E. (2015) Is science broken? *The Psychologist*. https://thepsychologist.bps.org.uk/volume-28/may-2015/science-broken  Accessed 16 October 2020.

Ricciato F., Wirthmann A., Giannakouris K., Reis F., and Skaliotis M. (2019) Trusted smart statistics: motivations and principles. To appear in *Statistical Journal of the International Association of Official Statistics.*

Rubin D.B. (1984) Bayesianly justifiable and relevant frequency calculations for the applies statistician*. The Annals of Statistics*, **12**, 1151-1172.

Sekhon H., Ennew C., Kharouf H., and Devlin J. (2014) Trustworthiness and trust: influences and implications. *Journal of Marketing Management,* **30(3-4)**, 409-430.

Singapore (2010) https://wcrif.org/guidance/singapore-statement  Accessed 12 October 2020.

Spanos A. (2010) Statistical adequacy and the trustworthiness of empirical evidence: statistical vs substantive information. *Economic Modelling*, **27**, 1436-1452.

Spiegelhalter D.J. (2017) Trust in numbers. *Journal of the Royal Statistical Society*, **180**, 949-965.

Trafimow D. (2014). Editorial. *Basic and Applied Social Psychology*, **36(1)**, 1–2.

Trafimow D. (2019) Five nonobvious changes in editorial practice for editors and rewiewers to consider when evaluating submissions in a post p<0.05 universe. *The American Statistician*, **73 suppl.1**, 340-345.

Trafimow D. and Marks M. (2015) Editorial, *Basic and Applied Social Psychology*, **37:1**, 1-2, DOI: 10.1080/01973533.2015.1012991

UKRI (2019) https://epsrc.ukri.org/files/funding/calls/2019/trustworthy-autonomous-systems-nodes-outline-call/  Accessed 16 October 2020.

UK Government (1998) *Statistics: A Matter of Trust*. https://assets.publishing.service.gov.uk/government/uploads/system/uploads/attachment_data/file/260823/report.pdf  Accessed 11th October 2020.

UKSA (2018) *Code of Practice*. https://code.statisticsauthority.gov.uk/wp-content/uploads/2018/02/Code-of-Practice-for-Statistics.pdf  Accessed 16 October 2020.

ValidateAI (2019) https://validateaiconference.com/  Accessed 16 October 2020.

Vichi M. and Hand D.J. (2019) Trusted smart statistics: the challenge of extracting usable aggregate information from new data sources. To appear in *Statistical Journal of the International Association of Official Statistics*.

Vincent J. (2020) Excel spreadsheet blamed for UK's 16,000 missing coronarvirus cases. https://www.theverge.com/2020/10/5/21502141/uk-missing-coronavirus-cases-excel-spreadsheet-error  Accessed 8[th] October 2020.

Wasserstein R.L., Schirm A.L., and Lazar N.A. (2019) Moving to a world beyond "p<0.05". *The American Statistician*, **73: sup 1**, 1-19.

Whyte K.P. and Crease R.P. (2010) Trust, expertise, and the philosophy of science. *Synthese*, **177**, 411-425